\documentclass[twocolumn,showpacs,preprintnumbers,amsmath,amssymb]{revtex4}

\usepackage{graphicx}
\usepackage{dcolumn}
\usepackage{bm}

\begin{document}

\title{Two-photon interference of multimode two-photon pairs 
with an unbalanced interferometer}

\author{Hayato Goto}
\author{Haibo Wang}
\author{Tomoyuki Horikiri}
\author{Yasuo Yanagihara}
\author{Takayoshi Kobayashi}
\affiliation{Core Research for Evolutional Science and Technology(CREST), 
Japan Science and Technology Corporation (JST)} 
\affiliation{Department of Physics, Graduate School of Science, 
University of Tokyo, 7-3-1 Hongo, Bunkyo, Tokyo, 113-0033, Japan} 

\date{\today}

\begin{abstract}
Two-photon interference of multimode two-photon pairs 
produced by an optical parametric oscillator has been 
observed for the first time with an unbalanced interferometer. 
The time correlation between the multimode two photons has 
a multi-peaked structure. 
This property of the multimode two-photon state induces 
two-photon interference depending on delay time. 
The nonclassicality of this interference is also discussed.
\end{abstract}


\maketitle

Quantum interference is one of the most interesting phenomena in 
quantum physics. 
Since the observation of nonclassical effects in the interference of 
two photons by Ghosh and Mandel \cite{Ghosh}, 
several types of quantum interference experiments have been demonstrated 
using correlated two-photon pairs 
generated by spontaneous parametric down-conversion (SPDC) 
\cite{Hong,Ou1989,Ou1990,OuFranson,Chiao,Rarity,Brendel,ShihMach}. 
In 
Refs. \cite{Ghosh,Hong,Ou1989,Ou1990,OuFranson,Chiao,Rarity,Brendel,ShihMach}, 
higher visibility of two-photon interference 
than in the classical case 
is discussed as a typical nonclassical effect. 
Another feature of quantum interference is characterized by 
the shorter period of interference than in the classical case. 
Fonseca \textit{et al.} have observed a twice narrower interference pattern 
than single-photon interference 
with correlated two-photon pairs generated by SPDC \cite{Fonseca}. 
Quantum lithography has been proposed by Boto \textit{et al.} 
in order to surpass 
classical diffraction limit utilizing this feature of quantum interference 
\cite{Boto}.
D'Angelo \textit{et al.} have reported 
a proof-of-principle quantum lithography \cite{ShihLithography}. 
This feature of quantum interference has also been 
confirmed using a conventional Mach-Zehnder interferometer \cite{Edamatsu}. 

\begin{figure}[htbp]
	\includegraphics[width=6.5cm, height=2.5cm]{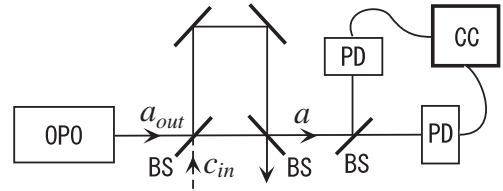}
	\caption{Block diagram of our experiment. 
						The output beam from an OPO is incident 
						at one of the input ports of an unbalanced interferometer. 
						The correlation function of one of the outputs of 
						the interferometer is observed with two photodetectors 
						and a coincidence counter. 
						BS: 50/50 beam splitter;
						PD: photodetector;
						CC: coincidence counter.}
	\label{sketch}
\end{figure}

In these quantum interference experiments, 
SPDC process has been used to prepare correlated two photons. 
Ou \textit{et al.} have succeeded in generating 
correlated two photons with new property 
by using an optical parametric oscillator (OPO) \cite{OuPRL,OuPRA}. 
The bandwidth of the two photons is narrow and 
the correlation time is long ($\sim$10 ns). 
This property of the two photons produced by an OPO has enabled to directly 
observe their correlation function by coincidence counting. 
These narrow-band two-photon state have been used to observe nonclassical 
photon statistics \cite{OuNonclassical}. 
Goto \textit{et al.} have recently reported the observation of 
another type of correlated two photons, 
that is, multimode two photons produced by an OPO \cite{Goto}.
The correlation function of the multimode two-photon pairs has 
a multi-peaked structure. 
In this Brief Report, 
we report the first observation of two-photon interference 
of the multimode two-photon pairs 
with an unbalanced interferometer. 
Our experiment is shown in Fig. \ref{sketch}. 
The output beam from an OPO 
is incident 
at one of the input ports of an unbalanced interferometer. 
The correlation function of one of the outputs of the interferometer 
is observed with two photodetectors and a coincidence counter. 
First, we discuss briefly what happens in this experiment. 
Next, we explain our experimental setup and show our experimental results. 
Finally, we discuss our results with theoretical calculation of 
the correlation function of the output from the interferometer. 
The nonclassical feature of the two-photon interference is also discussed.

The distance between multimode two photons produced by an OPO is 
$n\tau_rc$, where $\tau_r$ is the round-trip time of the OPO cavity, 
$c$ is the speed of light in the vacuum, 
and $n$ is an integer \cite{Goto}. 
We assume that 
the propagation time difference, $T$, between the short and long paths 
in the interferometer is nearly equal to $\tau_r /2$.  
There are two cases, Case1 and Case2: 
in Case 1, both the photons are reflected or transmitted 
at the first beam splitter of the interferometer; 
in Case 2, one of the two photons is reflected and the other is
transmitted there. 
The distance between two photons in the output of 
the interferometer is $n\tau_rc$ in Case 1 and 
$(n+1/2)\tau_rc$ in Case 2. 
Therefore, 
the two-photon pairs in Case 1 and Case 2 will induce 
the peaks of coincidence counts at delay times 
$n\tau_r$ and $(n+1/2)\tau_r$, respectively. 
This enables to distinguish Case 1 and Case 2 through delay time. 
The height of the peaks of coincidence counts 
at delay times $(n+1/2)\tau_r$ 
will be constant with regard to 
the path-length difference of the interferometer 
because two photons in Case 2 do not interfere with each other. 
On the other hand, in Case 1, 
the two-photon pairs provide two-photon interference 
because we can not say which path a two-photon pair propagate on. 
Therefore, the height of the peaks of coincidence counts 
at delay times $n\tau_r$ 
will change with regard to 
the path-length difference of the interferometer.
Thus, it is expected that 
two-photon interference depending on delay time will be observed.

\begin{figure}[htbp]
	\includegraphics[width=8.5cm, height=7.5cm]{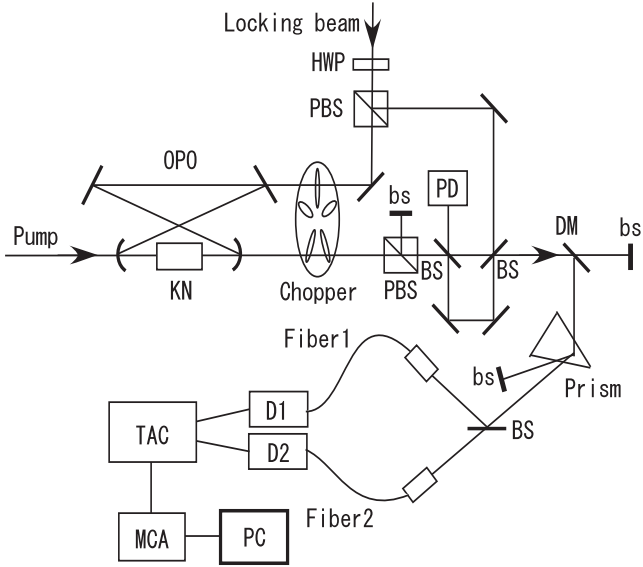}
	\caption{Schematic of the experimental setup. 
					PBS: polarization beam splitter; 
					HWP: half wave plate; 
					PD: photodetector for phase lock; 
					KN: KNbO$_3$ crystal; 
					DM: dichroic mirror; 
					BS: 50/50 beam splitter; 
					bs: beam stop; 
					D1 and D2: avalanche photodiodes; 
					TAC: time-to-amplitude converter; 
					MCA: multichannel analyzer.}
	\label{schematic}
\end{figure}

\begin{figure*}[htbp]
	\includegraphics[width=17cm, height=10.5cm]{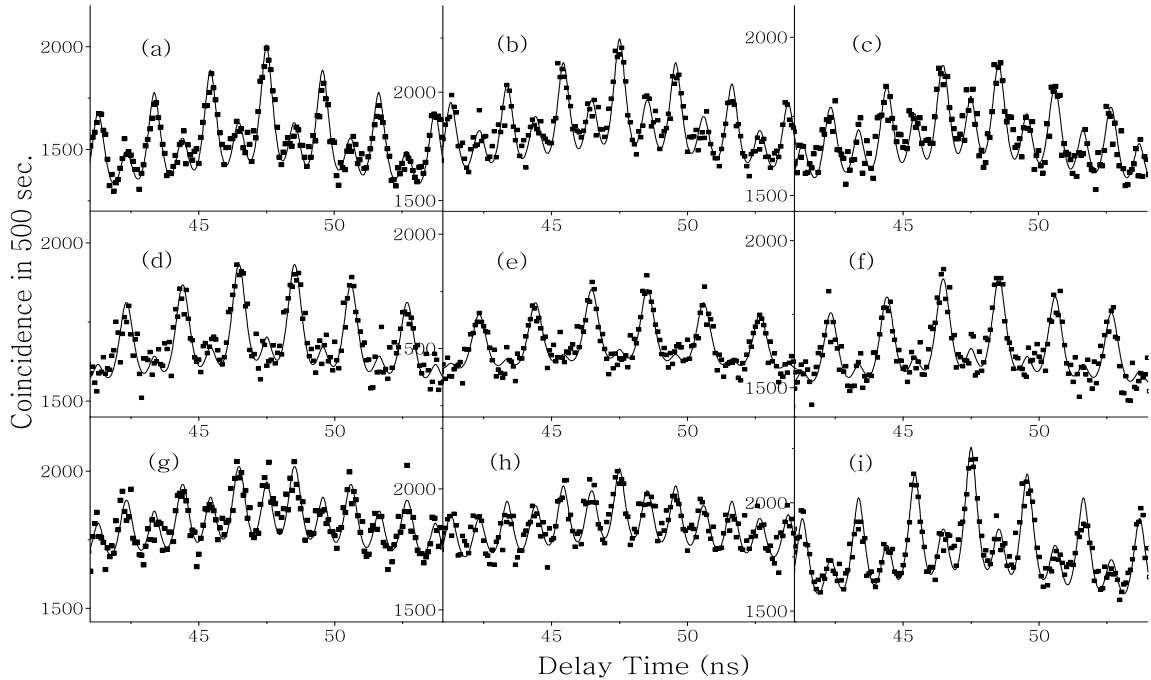}
	\caption{Experimental results. 
					The phase, $\theta$, 
					increases stepwise by $\pi /8$ from (a) to (i) 
					($\theta = (\pi/8) \times j$ ; $j=0,1,\cdots ,8$ ).
					The circles represent the measured coincidence counts.
					The lines are fits to Eq. (\ref{coincidence}), 
					where fitting parameters are 
					$C_1$, $C_2$, and $\theta$.
					The fitting result of $\theta$ is shown in Fig. \ref{result2}.
					The range of the data used for the fitting is from 18 ns to 77 ns.
					}
	\label{result1}
\end{figure*}

The schematic of the experimental setup is shown in Fig. \ref{schematic}. 
The differences between this setup and that used in our 
previous work \cite{Goto} 
are an unbalanced interferometer between the OPO and detectors, 
and a locking beam for the phase lock of the interferometer. 
The light source is 
a single-mode cw Ti:Sapphire laser of wavelength 860 nm. 
The round-trip length of the OPO is set long (560 mm) 
in order to time resolve the oscillatory structure 
in the correlation function. 
The output beam from the OPO is incident at one of the input ports 
of the interferometer. 
The path-length difference of the interferometer is 
set at about 29 cm, which gives $T\simeq 0.97$ ns. 
The phase difference of the interferometer is locked 
by a servo-control system. 
To prevent the locking beam for the interferometer 
from making noise, 
the beam propagates in the interferometer 
in the opposite direction to the signal beam from the OPO.
Furthermore, 
the polarization of the locking beam 
is perpendicular to that of the signal in order to remove the beam 
by a polarization beam splitter (see Fig. \ref{schematic}). 
One of the two outputs of the interferometer 
is split into two with a 50/50 beam splitter. 
The two beams are coupled to optical fibers 
and detected with avalanche photodiodes 
(APD, EG\&G SPCM-AQR-14). 
The coincidence counts of the signals from the two APDs are 
measured with a time-to-amplitude converter (TAC, ORTEC 567) and 
a multichannel analyzer (MCA, NAIG E-562). 
We measured the coincidence counts
at $\theta = (\pi /8) \times j$ ($j=0, 1, \cdots , 8$), 
where $\theta$ is the phase difference of the interferometer 
defined as the intensity of the output of the interferometer is 
proportional to $(1+\cos \theta )$ when classical light of wavelength 
860 nm is incident to the interferometer.
The experimental results at 
$\theta = (\pi /8) \times j$ ($j=0, 1, \cdots , 8$) are shown 
in Fig. \ref{result1}(a)-(i). 

From now, we discuss our experimental results. 
First, we derive the correlation function of the output from 
the interferometer. 
The output operator, $\tilde{a}(t)$, from one of the output ports of the 
interferometer
is expressed as 
\begin{align}
\tilde{a}(t) 
=& 
\frac{ \tilde{a}_{out} (t-T_S) + \tilde{c}_{in} (t-T_S)}{2} 
\nonumber
\\
&-
\frac{ \tilde{c}_{in}(t-T_L)- \tilde{a}_{out}(t-T_L)}{2}. 
\label{outputMZI}
\end{align}
Here $\tilde{a}(t)$ denotes a Fourier transform of a field operator, 
$a(\Omega )$, 
of frequency $\omega_0 + \Omega$ 
($\omega_0$ is the degenerate frequancy of the OPO). 
That is, it is defined as 
\begin{align}
\tilde{a}(t) = \frac{1}{\sqrt{2\pi}} 
\int d\Omega a(\Omega ) e^{-i(\omega_0 + \Omega )t}.
\end{align}
$a_{out}$ is the output operator of the OPO far below threshold 
\cite{Goto} and 
$c_{in}$ is an annihilation operator of 
the vacuum entering the interferometer from the other of the input ports 
(see Fig. \ref{sketch}). 
$cT_S$ and $cT_L$ are the short and long path lengths of the interferometer, 
respectively.
The intensity correlation function is derived as follows \cite{OuPRA,Goto}:
\begin{widetext}
\begin{align}
\Gamma (\tau ) 
&= 
\langle \tilde{a}^{\dagger} (t) \tilde{a}^{\dagger} (t+\tau ) 
\tilde{a}(t+\tau ) \tilde{a}(t) \rangle 
\nonumber
\\
&=
\left[
\frac{
2 \sqrt{\Gamma_{0} (\tau )} \cos \theta
+ \sqrt{\Gamma_{0} (\tau - T)} 
+ \sqrt{\Gamma_{0} (\tau + T)}}{4}
\right]^2
+
\left|
\frac{
2 \sqrt{\delta \Gamma_{0} (0)} 
+ e^{i\theta} \sqrt{\delta \Gamma_{0} (- T)} 
+ e^{-i\theta} \sqrt{\delta \Gamma_{0} (T)}}{4}
\right|^2
\nonumber \\
&+
\left|
\frac{
2 \sqrt{\delta \Gamma_{0} (\tau )} 
+ e^{i\theta} \sqrt{\delta \Gamma_{0} (\tau - T)} 
+ e^{-i\theta} \sqrt{\delta \Gamma_{0} (\tau + T)}}{4}
\right|^2,
\label{correlation}
\end{align}
\end{widetext}
with 
\begin{align}
\Gamma_0 (\tau )
&=
|\epsilon |^2 \left( \frac{F}{F_0} \right)^2 
e^{-\Omega_c |\tau | }
\frac{\sin^2 [(2N+1) \Omega_F \tau /2]}{\sin^2 (\Omega_F \tau /2)}
\label{correlation0}, 
\\
\delta
&=
\frac{4|\epsilon |^2}{\Omega_c^2}.
\end{align}
Here 
$\epsilon$ is the single-pass parametric amplitude gain; 
$F$ and $F_0$ are the finesse of the OPO with and without loss, 
respectively; 
$\Omega_c$ and $\Omega_F$ are the bandwidth and free spectral range 
of the OPO, respectively; 
$2N+1$ is the number of the longitudinal modes in the OPO output. 
$\Gamma_0(\tau )$ is a multi-peaked function of delay time $\tau$. 
The width of the peaks is about $\tau_r /(2N+1)$. 
We assume that $T$ is nearly equal to $\tau_r/2$ 
but the difference between $T$ and $\tau_r /2$ 
is longer 
than the width of the peaks. 
This assumption is feasible in our experiment. 
This results in the following approximations: 
\begin{align}
\Gamma_0 (\tau) \Gamma_0 (\tau \pm T) &\simeq 0, 
\Gamma_0 (\tau -T) \Gamma_0 (\tau +T) \simeq 0, 
\Gamma_0 (\pm T) \simeq 0. 
\end{align}
In addition, $\delta$ is much smaller than one when the OPO is operated 
far below threshold. 
Therefore, Eq. (\ref{correlation}) can be approximated as follows:
\begin{align}
\Gamma (\tau ) 
\simeq
\frac{
\Gamma_0 (\tau ) \cos^2 \theta
}{4}
+
\frac{
\Gamma_0 (\tau - T) 
+ \Gamma_0 (\tau + T)
}{16} + \frac{\delta \Gamma_0 (0)}{4}.
\label{correlationapp}
\end{align}
The first term in the right-hand side of Eq. (\ref{correlationapp}) 
corresponds to two-photon interference in Case 1. 
This is a periodic function of $\theta$ with a period $\pi$, 
which is a half of that of classical interference. 
It means that two-photon interference in Case 1 is nonclassical. 
The second term in the right-hand side of Eq. (\ref{correlationapp}), 
which is constant with regard to $\theta$, 
corresponds to coincidence counts in Case 2. 
These two terms are due to correlated two photons. 
The last term in the right-hand side of Eq. (\ref{correlationapp}) 
corresponds to the contribution from higher photon-number states than 
two. 
This term is not negligible when the pump power of the OPO 
is relatively high. 
As discussed in Ref. \cite{Goto}, 
the coincidence rate measured in experiments is 
an average of the correlation function 
over the resolving time, $T_R$, of detectors. 
According to Ref. \cite{Goto}, 
the coincidence counts measured in this experiment 
will become 
\begin{widetext}
\begin{align}
\Gamma_c (\tau ) 
=
C_1 
\left[
4 \Gamma_c^{(0)} (\tau ) \cos^2 \theta
+ \Gamma_c^{(0)} (\tau - T) 
+ \Gamma_c^{(0)} (\tau + T)
\right]
+
C_2, 
\label{coincidence}
\end{align}
with
\begin{align}
\Gamma_c^{(0)} (\tau )
=
e^{-\Omega_c |\tau -\tau_0|} 
\sum_n 
\left(
1+\frac{2 |\tau - n\tau_r -\tau_0| \ln 2}{T_R}
\right)
\exp
\left(
-
\frac{2 |\tau -n\tau_r -\tau_0| \ln 2}{T_R}
\right).
\end{align}
\end{widetext}
Here  $C_1$ and $C_2$ are constants and $\tau_0$ is an electric delay. 
The lines in Fig. \ref{result1} are fits to Eq. (\ref{coincidence}).
Fitting parameters are two constants, $C_1$ and $C_2$, 
and the phase difference $\theta$. 
Constant parameters are set as follows: 
$\tau_0=47.5$ ns; $\tau_r =2.07$ ns; $T_R =280$ ns; 
$\Omega_c /(2\pi ) =11$ MHz. 
The range of the data used for the fitting is from 18 ns to 77 ns, 
while the range plotted in Fig. \ref{result1} is 
from 41 ns to 54 ns. 
The fitting in Fig. \ref{result1} is good. 
The term, $C_2$, independent of the delay time, $\tau$, is mainly due to 
the last term in the right-hand side of Eq. (\ref{correlationapp}). 
In our experiment, $C_2$ is comparable to $C_1$. 
It means that the pump power of the OPO is relatively high 
and the contribution from higher photon-number states than two is 
not negligible. 
\begin{figure}[htbp]
	\includegraphics[width=8.5cm, height=7cm]{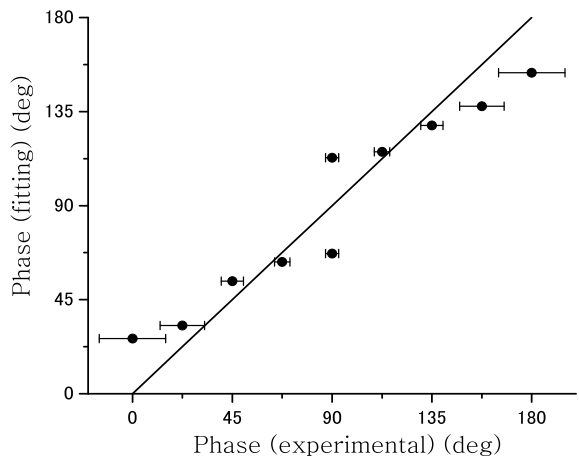}
	\caption{Phase(fitting) determined by the fitting 
					 shown in Fig. \ref{result1} is plotted against 
					 phase(experimental), which is the phase locked experimentally.
					 The inclination of the line is unity. 
					 The error bars are estimated from the fluctuation of the phase.}
	\label{result2}
\end{figure}
The phase determined from the fitting is plotted in Fig. \ref{result2} 
against the phase locked experimentally. 
The inclination of the line in Fig. \ref{result2} is unity. 
The error bars are estimated from the fluctuation of the phase. 
The deviations of the circles from the line are probably
due to the fluctuation of the phase difference, 
which is shown by error bars in Fig. \ref{result2}, and due to
an imperfect visibility, which makes larger the deviations around 
$\theta = 0$, $\pi /2$, and $\pi$. 
Taking these points into consideration, 
the correspondence between theory and experiment 
seems not to be bad. 
From these fitting results, 
it is concluded that 
our experimental results can be explained by Eq. (\ref{coincidence}), 
which is derived from Eq. (\ref{correlationapp}). 
Since Eq. (\ref{correlationapp}) includes a term corresponding to 
nonclassical interference, 
it is also concluded that 
nonclassicality has appeared in this experimental results.

In conclusion, we have observed two-photon interference of 
multimode two-photon pairs produced by an OPO 
for the first time 
with an unbalanced interferometer. 
This two-photon interference has been dependent on delay time. 
The experimental results have been explained theoretically. 
Nonclassical feature of the two-photon interference 
has been also discussed.

\end{document}